\documentclass[aps,pra,twocolumn,eqsecnum,superscriptaddress,showpacs,floatfix,nofootinbib]{revtex4}

\usepackage{psfrag,graphicx}
\usepackage{dcolumn}
\usepackage{amsmath,amssymb}
\usepackage{bm}

\newcommand{\be}{\begin{equation}}
\newcommand{\ee}{\end{equation}}
\newcommand{\bq}{\begin{eqnarray}}
\newcommand{\eq}{\end{eqnarray}}

\newcommand{\no}{\nonumber\\}

\def\be{\begin{equation}}
\def\ee{\end{equation}}
\def\bea{\begin{eqnarray}}
\def\eea{\end{eqnarray}}

\def\1/2{\frac{1}{2}}

\begin{document}

\title{ Effective three-body interactions in triangular optical lattices}

\author{Jiannis K. Pachos}
\affiliation{Department of Applied Mathematics and Theoretical
Physics, University of Cambridge, Cambridge CB3 0WA, UK,}
\author{Enrique Rico}
\affiliation{Department d'Estructura i Constituents
de la Mat\` eria, Universitat de Barcelona, 08028, Barcelona, Spain.}

\date{\today}

\begin{abstract}

We demonstrate that a triangular optical lattice of two atomic
species, bosonic or fermionic, can be employed to generate a
variety of novel spin-$1/2$ Hamiltonians. These include effective
three-spin interactions resulting from the possibility of atoms
tunneling along two different paths. Such interactions can be
employed to simulate particular one or two dimensional physical
systems with ground states that possess a rich structure and undergo a
variety of quantum phase transitions.
In addition, tunneling can be activated by employing Raman transitions,
thus creating an effective Hamiltonian that does not preserve the number
of atoms of each species. In the presence of external electromagnetic fields,
resulting in complex tunneling couplings, we obtain effective
Hamiltonians that break chiral symmetry.
The ground states of these Hamiltonians can be used for the physical
implementation of geometrical or topological objects.

\end{abstract}


\maketitle

\section{Introduction}

With the development of optical lattice technology
\cite{Raithel,Mandel1,Mandel3}, considerable attention has been
focused on the experimental simulation of a variety of
many-particle systems, such as spin chains
\cite{Jaks98,Kukl,Jask03,Duan}. This provides the possibility to
probe and realise complex quantum models with unique properties in the
laboratory. Such examples, that are of interest in
various areas of physics, are the systems that include many-body
interactions. The latter have been hard to study in
the past due to the difficulty in controlling them externally and
isolating them from the environment \cite{Mizel}. To overcome these
problems, techniques have been developed in quantum optics
\cite{Cirac1,Carl,Roberts} which minimise imperfections and
impurities in the implementation of the desired structures, thus
paving the way for the consideration of such ``higher order''
phenomena of multi-particle interactions. Their applications could be
of much interest to cold atom technology as well as to condensed
matter physics and quantum information.

In this paper we obtain the interaction terms of bosonic or
fermionic lattices of two species of atoms, denoted here by
$\uparrow$ and $\downarrow$ (see \cite{Jaks98,Kukl,Duan}). These
can be two different hyperfine ground states of the same atom coupled
via an excited state by a Raman transition. The system is brought
initially into the Mott insulator phase where the number of atoms at
each site of the lattice is well defined. By restricting to the
case of only one atom per site it is possible to characterise the
system by pseudo-spin basis states provided by internal ground
states of the atom. Interactions between atoms in different sites
are facilitated by virtual transitions. These are dictated by the
tunneling coupling $J$ from one site to its neighbours and by
collisional couplings $U$ that take place when two or more atoms
are within the same site.

In the following we consider the case of weak tunneling
couplings, $J\ll U$, assuring that we are always in the Mott
insulator regime. Our aim is to construct a perturbative study of the effective
interactions with respect to the small parameter $J/U$. Up to the
third order this expansion will provide Hamiltonians that include
three-spin interactions. These multi-particle interactions can be,
in principle, realised with the near future technology. The main
physical requirement is large collisional couplings, $U$, which
can be obtained experimentally by Feshbach resonances
\cite{Inouye,Donley,Kokkelmans}. First theoretical \cite{Mies} and
experimental \cite{Donley1} advances are already promising. Hence,
the time interval needed for those higher order terms to have a
significant effect can be well within the coherence times of the
system.

Several applications spring out from our studies. The systematic
description of the low energy Hamiltonian provides the means for
the advanced control of the three-spin interactions simulated in
the lattice. Hence, different physical models can be realised, with
ground states that present a rich structure such as multiple
degeneracies and a variety of quantum phase transitions
\cite{Sachdev,Sachdev1,Pachos04}. Some of these multi-spin
interactions have been theoretically studied in the past in
the context of the hard rod boson \cite{Pens88,Iglo89,Bass,Fend},
using self-duality symmetries \cite{Turb,Pens82}. Phase transitions
between the corresponding ground states have been
analysed \cite{Iglo87,Chris}. Subsequently, these phases may
also be viewed as possible phases of the initial system, that is in
the Mott insulator, where the behaviour of its ground state can be
controlled at will \cite{Laughlin}.

The paper is organised as follows. In Section
\ref{lattice}, we present the physical system and the conditions
required to obtain three-body interactions. The effective three-spin
Hamiltonians for the case of bosonic or fermionic species of
atoms in a system of three sites on a lattice are given in Sec.
\ref{eff}. In Sec. \ref{Raman} we study the
effect Raman transitions can have on the tunneling process and
generalised effective Hamiltonians are presented that
do not preserve the number of atoms of each species. These are of
particular interest for the construction of certain geometrical
evolutions. In Sec. \ref{complex} complex tunnelings are
considered and the generation of chiral ground states is
presented. In Sec. \ref{onedim} our results are extended toward
the construction of one dimensional models and several
applications are discussed. In Sec. \ref{conclusions} we present an
outlook and the conclusions. Finally, in
the Appendix, two alternative methods are presented for the
perturbation theory that results in the three-spin interactions.

\section{The Physical Model}
\label{lattice}

\begin{center}
\begin{figure}[!h]
\resizebox{!}{2.9cm} {\includegraphics{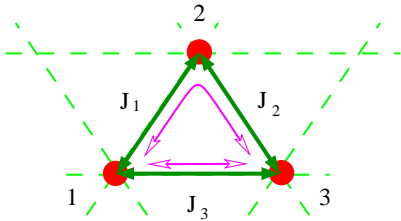} }
\caption{\label{triangle} The basic building block for the triangular
    lattice configuration. Three-spin
    interaction terms appear between sites $1$, $2$ and $3$. For
    example, tunneling between $1$ and $3$ can happen through two
    different paths, directly and through site $2$. The latter
    results in an exchange interaction between $1$ and $3$ that is
    influenced by the state of site $2$.}
\end{figure}
\end{center}

Let us consider a cloud of ultra cold neutral atoms superimposed
with several optical lattices
\cite{Jaks98,Kukl,Jask03,Duan,Pachos03}. For sufficiently strong
intensities of the laser field this system can be placed in the
Mott insulator phase where the expectation value of only one
particle per lattice site is energetically allowed \cite{Mandel3}.
We are interested in the particular setup of lattices that form an
equilateral triangular configuration, as shown
in Fig. \ref{triangle}. This allows for the simultaneous
superposition of the positional wave functions of the atoms
belonging to the three sites. As we shall see in the following this
results in the generation of three-spin interaction.

The main contributions to the dynamics of the atoms in the lattice
sites are given by the collisions of the atoms within the same
site and the tunneling transitions of the atoms between
neighbouring sites. In particular, the coupling of the collisional
interaction for atoms in the same site are taken to be very large
in magnitude, while they are supposed to vanish when they are in
different sites. Due to the low temperature of the system, this
term is completely characterised by the s-wave scattering length.
Furthermore, the overlap of the Wannier wave functions between
adjacent sites determines the tunneling amplitude, $J$, of the
atoms from one site to its neighbours. Here, the relative rate
between the tunneling and the collisional interaction term is supposed
to be very small, i.e. $J\ll U$, so that the state of the system is
mainly dominated by the collisional interaction.

The Hamiltonian describing the three lattice sites with three
atoms of species $\sigma=\{\uparrow,\downarrow\}$ subject to the
above interactions is given by
\begin{equation}
H=H^{(0)}+V,
\label{ham}
\end{equation}
with
\begin{equation}
\begin{split}
H^{(0)}&=\frac{1}{2} \sum _{j \sigma \sigma'} U_{\sigma \sigma'}
: n_{j\sigma} n_{j\sigma'}:, \\
V&=-\sum_{j\sigma} (J^\sigma_{j} a_{j\sigma}^\dagger a_{j+1
\sigma} +\text{H.c.}),
\end{split}
\nonumber
\end{equation}
where $a_{j \sigma}$ denotes the annihilation operator of atoms of
species $\sigma$ at site $j$. The annihilation operator can
describe fermions or bosons satisfying commutation or
anticommutation relations respectively given by
\begin{equation}
\begin{split}
&[a_{j \sigma},a_{j'\sigma'}^\dagger]_{\pm}=\delta_{jj'}
\delta_{\sigma \sigma' }, \\
&[a_{j\sigma},a_{j'\sigma'}]_{\pm}=[a_{j\sigma}^\dagger
,a_{j'\sigma'}^\dagger]_{\pm}=0,
\end{split}
\label{comm}
\end{equation}
where the $\pm$ sign denotes the anticommutator or the commutator.
The operator $n_{j\sigma}$ is the corresponding number operator
and $:...:$ denotes normal ordering of the product of the creation
and annihilation operators. The Hamiltonian $H^{(0)}$ is the
lowest order in the expansion with respect to the tunneling
interaction.

Due to the large collisional couplings activated when two or more
atoms are present within the same site, the weak tunneling
transitions do not change the average number of atoms per site.
This is achieved by adiabatic elimination of higher population
states along the evolution leading eventually to an effective
Hamiltonian (see Appendix). The latter allows virtual transitions
between these levels providing eventually non-trivial evolutions. As
we shall see in the Appendix it is possible to describe the low energy
evolution of the bosonic or fermionic system up to the third order
in the tunneling interaction by the effective Hamiltonian
\begin{equation}
H_{\text{eff}}=-\sum _\gamma {V_{\alpha \gamma} V_{\gamma \beta}
\over
  E_\gamma} +  \sum _{\gamma \delta} {V_{\alpha \gamma} V_{\gamma
    \delta} V_{\delta \beta} \over E_\gamma E_\delta}.
\label{effective}
\end{equation}
The indices $\alpha$, $\beta$ refer to states with one atom
per site while $\gamma$, $\delta$ refer to states with two or more
atomic populations per site, $E_\gamma$ are the eigenvalues of the
collisional part, $H^{(0)}$, while we neglected fast rotating
terms effective for long time intervals.

It is instructive to estimate the energy scales involved in such a
physical system. We would like to have a significant effect of the
three-spin interaction within the decoherence
times of the experimental system, which we can take here to be of
the order of several 10 ms. It is possible to vary the tunneling
interactions from zero to some maximum value which we can take
here to be of the order of $J/\hbar\sim$1 kHz \cite{Mandel1}. In
order to have a significant effect from the term $J^3/U^2$
produced within the decoherence time one should choose $U/ \hbar
\sim$10 kHz. This can be achieved experimentally by moving close to a
Feshbach resonance \cite{Inouye,Donley,Kokkelmans,Bolda}, where
$U$ can take significantly large values as long as trap losses,
attributed to three body collisions or production of molecules, remain
negligible. With respect to these parameters we have $(J/U)^2\sim
10^{-2}$, which is within the Mott insulator
regime, while the next order in perturbation theory is an order of
magnitude smaller than the one considered here and hence
negligible. Note however, that new interaction terms arise only
in $5^{th}$ order in perturbation due to the triangular geometry of
the optical lattice. This places the requirements of our proposal for
detecting the effect of three-spin interactions within the range
of the possible experimental values of the near future
technology.

Within the regime of single atom occupancy per site it is possible
to switch to the pseudo-spin basis of states of the site $j$ given
by $| \! \uparrow\rangle\equiv |n_{j\uparrow}=
1,n_{j\downarrow}=0\rangle$ and $|\! \downarrow\rangle \equiv
|n_{j\uparrow}= 0,n_{j\downarrow} = 1\rangle$. Hence, the
effective Hamiltonian can be given in terms of Pauli
matrices acting on states expressed in the pseudo-spin basis. The
symmetries of the initial Hamiltonian $H$ restrict to a large
degree the form of the low energy expansion. For example, the
conservation of the atom number of each species corresponds, in
the spin basis, to the conservation of the total $z$-spin. Hence,
any rotation on the $xy$-spin plane leaves the Hamiltonian
invariant. This fact limits the possible spin operators that can
contribute to the effective low energy interactions. Possible
terms of the effective Hamiltonian are given by $\{ \sigma^z_{j} \}$
for the one-body interaction, $\{\sigma^z_{j}
\sigma^z_{j+1},\,\sigma^x_{j} \sigma^x_{j+1} +\sigma^y_{j}
\sigma^y_{j+1} \}$ for the two body interaction or $\{\sigma^z_{j}
\sigma^z_{j+1}\sigma^z_{j+2},\,(\sigma^x_{j} \sigma^x_{j+1}
+\sigma^y_{j} \sigma^y_{j+1})\sigma^z_{j+2} \}$ for the three-body
interactions where $\sigma_4=\sigma_1$, (see Fig.
\ref{triangle}). As we can easily verify, the three-spin operators
break parity symmetry, which is explicitly given by the
transformation $\uparrow \leftrightarrow \downarrow$. This
indicates that their coupling coefficient should also be
asymmetric with respect to this transformation, as the original
atomic system possesses this symmetry. Indeed, in the next section
we shall see how these terms are generated in the optical lattice
setup.

Another important insight for the ground states of the presented
Hamiltonians comes from the geometry of the lattice. In the case
considered here,
the triangular pattern allows for the generation of exotic ground
states due to frustration, that is ground states that are not
minimising the energy of the individual Hamiltonians of each link
of the triangle. This effectively allows for the presence of
multiple degeneracy in the ground state of the system as we shall
see in particular examples.

\section{The effective three-spin Interactions}
\label{eff}

\subsection{The bosonic model}

\begin{center}
\begin{figure}[!ht]
\resizebox{!}{7.0cm}{\includegraphics{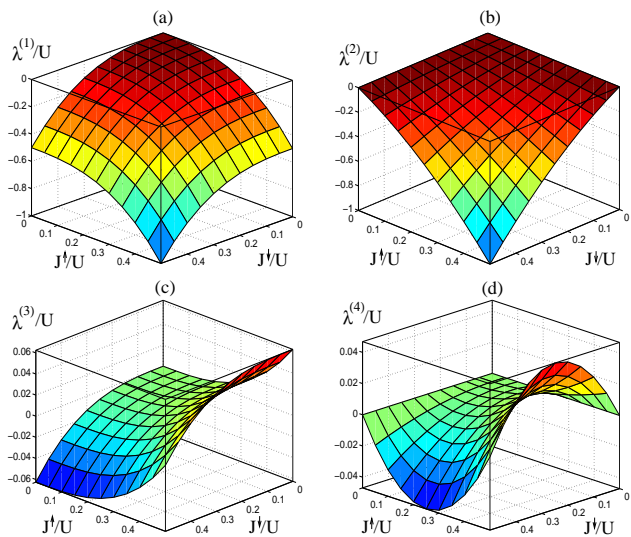}}
\caption{\label{comb1}The effective couplings (a) $\lambda^{(1)}$, (b)
  $\lambda^{(2)}$, (c) $\lambda^{(3)}$ and (d) $\lambda^{(4)}$ as
  functions of the
  tunneling couplings $J^{\uparrow}/U$ and $J^{\downarrow}/U$, where
  we have set the tunneling couplings to be $J^\sigma_{1}=J^\sigma_{2}
  =J^\sigma_{3}$ and the collisional couplings to be $U_{ \uparrow \uparrow
  }=U_{\uparrow \downarrow }=U_{\downarrow \downarrow }=U$. All the
  parameters are normalised with respect to $U$.}
\end{figure}
\end{center}

Consider the low energy evolution of the triangular system given
in Fig. \ref{triangle} of three atoms in three sites of the
lattice without the application of any external field. Different
rates in the tunneling parameter can then be achieved by tuning the
intensities of the laser field corresponding to the different
directions of the triangle. By applying the perturbative expansion
(\ref{effective}) up to the third order we obtain that
the system can effectively be
described by
\begin{equation}
\label{ham1}
\begin{split}
H_{\text{eff}} &=\sum_{j=1}^3 \Big[A_j \mathbb{I} +  B_j
\sigma^z_j +
\\ &\lambda^{(1)}_j
\sigma^z_j \sigma^z_{j+1} + \lambda^{(2)}_j (\sigma^x_{j} \sigma^x_{j+1}
+\sigma^y_{j} \sigma^y_{j+1}) +
  \\ \\ &\lambda^{(3)} \sigma^z_{j} \sigma^z_{j+1}
  \sigma^z_{j+2}+\lambda^{(4)}_j (\sigma^x_{j} \sigma^z_{j+1}
  \sigma^x_{j+2}+ \sigma^y_{j} \sigma^z_{j+1} \sigma^y_{j+2})
  \Big],
\end{split}
\end{equation}
where $\sigma_{j}^{\alpha}$ is the $\alpha$ Pauli matrix with the
usual commutation properties
$[\sigma_{j}^{\nu},\sigma_{k}^{\mu}]=2i \epsilon^{\nu \mu \omega}
\delta_{jk} \sigma_{j}^{\omega}$. The three-spin interactions
presented in the last line can be viewed as the two spin
interactions of the second line controlled by the third spin
(being spin up or down) through an additional $\sigma^z$ operator.
The couplings, $A$, $B$, and $\lambda^{(i)}$, are given as
expansions in ${J^\sigma}/U_{\sigma \sigma'}$ by
\begin{equation}
\begin{split}
A_j=&-J_{1}^{\uparrow}J_{2}^{\uparrow}J_{3}^{\uparrow} \big(
\frac{3}{2U_{\uparrow \uparrow}^2}+ \frac{1}{2U_{\uparrow
\downarrow}^2} + \frac{1}{U_{\uparrow \downarrow}U_{ \uparrow
\uparrow }} \big)-
\\ & {J_{j}^{\uparrow}}^2\big(\frac{1}{ U_{\uparrow \uparrow}}+
\frac{1}{2U_{\uparrow \downarrow}}
\big) + (\uparrow \leftrightarrow
\downarrow), \\
B_j= & -\frac{{J_{j}^{\uparrow}}^2 +{J_{j+2}^{\uparrow}}^2}{U_{\uparrow
      \uparrow}}-\frac{J_{1}^{\uparrow} J_{2}^{\uparrow}
      J_{3}^{\uparrow}}{U_{\uparrow \uparrow}}\big( \frac{1}{U_{\uparrow
      \downarrow}}+\frac{9}{2 U_{\uparrow \uparrow}}\big)-\\
      &( \uparrow\leftrightarrow \downarrow),
\nonumber
\end{split}
\end{equation}
\vspace{-0.5cm}
\begin{equation}
\begin{split}
\lambda^{(1)}_j =&-J_{1}^{\uparrow}J_{2}^{\uparrow}J_{3}^{\uparrow}
\big(\frac{9}{2U_{\uparrow \uparrow}^2}- \frac{1}{2U_{\uparrow
\downarrow}^2} - \frac{1}{U_{\uparrow \downarrow }U_{ \uparrow
\uparrow}} \big) -\\
& {J^{\uparrow}_{j}}^2 \big( \frac{1}{U_{\uparrow \uparrow}}
-\frac{1}{2U_{\uparrow \downarrow }}\big) + (\uparrow
\leftrightarrow
\downarrow), \\
\lambda^{(2)}_j=& - J_{j}^{\downarrow} J_{j+1}^{\uparrow}
J_{j+2}^{\uparrow} \big( \frac{3}{2U_{\uparrow
\downarrow}^2}+\frac{1}{2 U_{\uparrow \uparrow}^2}+
\frac{1}{U_{\uparrow \downarrow} U_{\uparrow \uparrow }} \big )-\\
& \frac{J^{\uparrow}_{j} J^{\downarrow}_{j}}{2 U_{\uparrow
    \downarrow}} + (\uparrow \leftrightarrow \downarrow),\\
\lambda^{(3)}=& -\frac{J_{1}^{\uparrow} J_{2}^{\uparrow}
  J_{3}^{\uparrow}}{U_{ \uparrow \uparrow }} \big( \frac{3}{2U_{
  \uparrow \uparrow}}-\frac{1}{U_{ \uparrow \downarrow} }\big)-
  (\uparrow \leftrightarrow \downarrow) ,\\
\lambda^{(4)}_j=&-\frac{J_{j}^{\uparrow} J_{j+1}^{\uparrow}
  J_{j+2}^{\downarrow}}{U_{ \uparrow \uparrow }} \big(\frac{1}{2 U_{
  \uparrow \uparrow}}+ \frac{1}{U_{\uparrow \downarrow }} \big ) -
  (\uparrow \leftrightarrow \downarrow),
\label{effcouplings}
\end{split}
\end{equation}
where the symbol $(\uparrow \leftrightarrow \downarrow)$
denotes the repeating of the same term as on its left, but with
the $\uparrow$ and $\downarrow$ indices interchanged. The $A$ term
contributes to an overall phase factor in the time evolution of the
system and can be ignored. The $B$ term
can easily be eliminated and an arbitrary magnetic field term of
the form $\sum_j \vec{B}\cdot \vec{\sigma}$ can be added by
applying a Raman transition with the appropriate laser fields. The
behaviour of the effective couplings as functions of the tunneling
and collisional couplings is given in Fig. \ref{comb1}.

\begin{center}
\begin{figure}[!ht]
\resizebox{!}{6.0cm}{\includegraphics{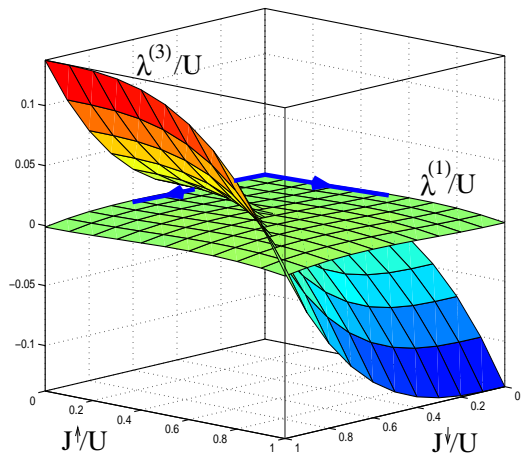}}
\caption{\label{zzz} The effective couplings $\lambda^{(1)}$ and
  $\lambda^{(3)}$
  are plotted against $J^{\uparrow}/U$ and $J^{\downarrow}/U$ for $U_{
  \uparrow \uparrow }=U_{\downarrow \downarrow }=2.12U$ and
  $U_{\uparrow \downarrow }=U$. The coupling $\lambda^{(1)}$ appears almost
  constant and zero as the unequal collisional terms can create a
  plateau area for a small range of the tunneling couplings, while
  $\lambda^{(3)}$ can be varied freely to positive or negative values.}
\end{figure}
\end{center}

One can isolate different parts of Hamiltonian (\ref{ham1}), each
one including a three-spin interaction term, by varying the
tunneling and/or the collisional couplings appropriately so that
particular terms in (\ref{ham1}) vanish, while others are freely
varied. An example of this can be seen in Fig. \ref{zzz} where the
couplings $\lambda^{(1)}$ and $\lambda^{(3)}$ are depicted. There,
for the special choice of the collisional terms, $ U_{
  \uparrow \uparrow }=U_{\downarrow \downarrow }=2.12U_{\uparrow
  \downarrow }$, the $\lambda^{(1)}$
coupling is kept to zero for a wide range of the tunneling
couplings, while the three-spin coupling, $\lambda^{(3)}$, can take
any arbitrary value. One can also suppress the exchange
interactions by keeping one of the two tunneling couplings zero
without affecting the freedom in obtaining arbitrary positive or
negative values for $\lambda^{(3)}$ as seen in Fig. \ref{zzz}.
Hence, the one dimensional Hamiltonian of the form
\begin{equation}
H(B_x,B_z)= - \sum_j \big(B_x \sigma_j^x +B_z \sigma_j^z +
\sigma_j^z \sigma_{j+1}^z \sigma_{j+2}^z\big) \nonumber
\end{equation}
can be simulated in the optical lattice where all of its couplings
can be arbitrarily and independently varied. The three-spin
interaction term of this Hamiltonian possesses fourfold degeneracy
in its ground state, spanned by the states $\{|\uparrow \uparrow
\uparrow\rangle,|\uparrow \downarrow \downarrow\rangle,|\downarrow
\uparrow \downarrow\rangle,|\downarrow \downarrow \uparrow\rangle
\}$. The criticality behaviour of this model has been extensively
studied in the past \cite{Pens88,Chris}, where it is shown to
present first and second order phase transitions. In particular,
for $B_z=0$ its self dual character can be demonstrated
\cite{Turb,Pens82}. To explicitly show that let us define the dual
operators
\begin{equation}
\bar \sigma^x_j \equiv \sigma^z_j \sigma^z_{j+1} \sigma^z_{j+2}
,\,\,\,\, \bar \sigma^z_j \equiv \prod_{k =0} ^\infty
\sigma^x_{i-3k} \sigma^x_{i-3k-1}, \nonumber
\end{equation}
that also satisfy the usual Pauli spin algebra. We can re-express
the Hamiltonian $H(B_x,0)$ with respect to the dual operators
obtaining finally
\begin{equation}
H(B_x,0)= B_xH(B_x^{-1},0). \nonumber
\end{equation}
This equation of self duality indicates that if there is one
critical point then it should be at $|B_x|= 1$ as has also been
verified numerically. Furthermore, the two spin interactions $
\sigma^z_j \sigma^z_{j+1}$ has a degeneracy with a $Z_2$ symmetry
while $\sigma^z_j \sigma^z_{j+1}\sigma^z_{j+2}$ has a three fold
degeneracy leading to a $Z_3$ symmetry. By varying the corresponding
couplings of the effective Hamiltonian it is possible to induce
transitions to and from the $Z_2 $ and $Z_3$ ordered phases in a
similar fashion as has been theoretically demonstrated in \cite{Sachdev1}.

\subsection{The fermionic model}

\begin{center}
\begin{figure}[!ht]
\resizebox{!}{7.0cm}{\includegraphics{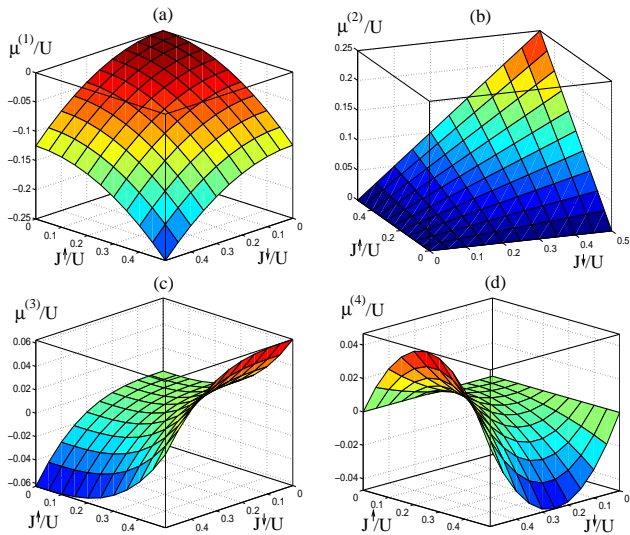}}
\caption{\label{comb2}The effective couplings (a) $\mu^{(1)}$, (b)
  $\mu^{(2)}$, (c) $\mu^{(3)}$ and (d) $\mu^{(4)}$ as functions of the
  tunneling couplings $J^{\uparrow}/U$ and $J^{\downarrow}/U$, where
  we have set the tunneling couplings to be $J^\sigma_{1}=J^\sigma_{2}
  =J^\sigma_{3}$.}
\end{figure}
\end{center}

Alternatively, one can consider the case of fermionic atoms and
derive the effective interactions they induce up to the third
order. Compared to the couplings in the bosonic case we now have
$U_{\uparrow \downarrow }=U$ being the only one that is present. The
Pauli exclusion principle can be signalled by having $U_{ \uparrow
\uparrow },U_{\downarrow \downarrow}\rightarrow \infty$ that
eventually forbids two fermionic atoms of the same
species from occupying the same site. Keeping terms up
to the third order in $J_{j}^{\sigma} /U$ and employing the
anticommutation relations (\ref{comm}) we obtain the effective
Hamiltonian
\begin{equation}
\begin{split}
&H_{\text{eff}}=\sum_{j=1}^3 \Big[ \mu^{(1)}_j (\mathbb{I}-
\sigma_{j}^z \sigma_{j+1}^z) +\mu^{(3)}
(\sigma_{j}^z -\sigma_{1}^z \sigma_{2}^z \sigma_{3}^z)+\\
&\mu^{(2)}_{j} (\sigma_{j}^x \sigma_{j+1}^x \!\!+\!\sigma_{j}^y
\sigma_{j+1}^y) \!+\!\mu^{(4)}_j (\sigma_{j}^x \sigma_{j+1}^z
\sigma_{j+2}^x \!\!+\! \sigma_{j}^y \sigma_{j+1}^z \sigma_{j+2}^y
) \Big],
\end{split}
\nonumber
\end{equation}
where the effective couplings are a function of the initial
variables of the Hamiltonian (\ref{ham}) and
\begin{equation}
\begin{split}
&\mu^{(1)}_j=
  -\frac{1}{2U}({J_{j}^{\uparrow}}^2+{J_{j}^{\downarrow}}^2),\,\,\,\,\,
\mu^{(2)}_{j}=\frac{1}{U}J_{j}^{\uparrow}J_{j}^{\downarrow},\\
&\mu^{(3)}=-\frac{1 }{2U^2}(J_{1}^{\uparrow}
      J_{2}^{\uparrow} J_{3}^{\uparrow}-J_{1}^{\downarrow}
      J_{2}^{\downarrow} J_{3}^{\downarrow} ),\\
&\mu^{(4)}_j=\frac{3 }{2U^2}(J_{j}^{\uparrow}
      J_{j+1}^{\uparrow} J_{j+2}^{\downarrow}-J_{j}^{\downarrow}
      J_{j+1}^{\downarrow} J_{j+2}^{\uparrow} ).
\end{split}
\nonumber
\end{equation}
The dependence of the coupling terms on the parameters of the
initial Hamiltonian is simpler than in the bosonic case. Nevertheless,
they can express a similar behaviour as can been seen in
Fig. \ref{comb2}.
If the tunneling constants do not depend
on the pseudo-spin orientation then any three-spin interaction
vanishes. Nevertheless, when the tunneling amplitudes depend on
the spin and by having just one of the orientation with non-zero
tunneling, then just the diagonal two- and three-spin interactions
remain.

\section{Raman activated tunnelings}\label{Raman}

A number of variations of the previous Hamiltonians are possible
by employing techniques available from quantum optics
\cite{Jaks98,Duan}. An interesting example involves the
application of Raman transitions during the tunneling process. These
transitions involve the direct coupling of the two atomic states
$\uparrow$ and $\downarrow$. Consequently they are not atom-number
preserving for each of the species.

\subsection{The general case}

Let us first consider the case where the
atoms are strongly trapped by an optical lattice as in the previous
sections. If the lasers producing the Raman transition are
forming standing waves it is possible to activate tunneling
transitions of atoms that simultaneously experience a change in their
internal state. As we shall see in the following the resulting
Hamiltonian is given by an $SU(2)$ rotation applied to each Pauli
matrix of the Hamiltonian (\ref{ham1}).

In particular, we shall consider the case of activating the tunneling with the
application of two individual Raman transitions. These transitions consist of
four paired laser beams $L_1$, $L_2$ and $L_1'$, $L_2'$, each pair having a blue
detuning $\Delta$ and $\Delta'$, different for the two different transitions. The
phases and amplitudes of the laser beams can be properly tuned so that the first
Raman transition allows the tunneling of the state
\begin{equation}
|+\rangle \equiv \cos \theta |a\rangle + \sin \theta e
  ^{-i\phi } |b\rangle
\nonumber
\end{equation}
with tunneling rate $J_+$ between two neighbouring sites, while the
second one activates the tunneling of the state
\begin{equation}
|-\rangle \equiv \sin \theta |a\rangle - \cos \theta e
  ^{-i\phi } |b\rangle,
\nonumber
\end{equation}
by an additional phase difference of $\pi$ between the lasers
$L_1'$, $L_2'$, with an effective tunneling rate $J_-$. In the
above equations $\phi$ denotes the  phase difference between the
$L_i$ laser field, while $\tan \theta =|\Omega_2 /\Omega_1|$,
where $\Omega_i$ are their corresponding Rabi frequencies. Hence,
the effective tunneling term is given by
\begin{equation}
V_c=-\sum_i (J_+ {c^+_i}^\dagger c^+_{i+1} +J_-{c^-_i}^\dagger
c^-_{i+1} +\text{H.c.}), \nonumber
\end{equation}
where the tunneling couplings $J_+ $ and $J_-$ are given by the
potential barrier of the initial optical lattice superposed by the
potential reduction due to the Raman transitions In addition, the
creation and annihilation operators are given as an $SU(2)$
rotation of the initial ones, i.e.
\[
\left( \begin{array}{c}
c^+_i \\
c^-_i
        \end{array} \right)
=g(\phi,\theta) \left( \begin{array}{c} a_i
\\
b_i
      \end{array} \right)
\]
with the unitary $SU(2)$ matrix
\[
g(\phi,\theta) = \left( \begin{array}{cc} \cos \theta & e^{i \phi}
\sin \theta
\\
\sin \theta & -e^{i\phi} \cos \theta
      \end{array} \right) .
\]
Hence, the resulting tunneling Hamiltonian can be obtained from the
initial one via an $SU(2)$ rotation $V_c= gVg^\dagger$, where the
corresponding tunneling couplings are formally identified, i.e.
$J^+=J^\uparrow$ and $J^-=J^\downarrow$. Note that the collisional
Hamiltonian is not affected by the Raman transitions, and hence
it is not transformed under $g$ rotations.

It is easy to derive the effective Hamiltonian for this
transformation using the perturbative expansion. Indeed, from
expressions (\ref{second}) and (\ref{third}) of the Appendix we
straightforwardly obtain the second and third order terms of the
Hamiltonian $\tilde H_{\text{eff}}$ that appear after the
application of the Raman transition. They are given by an $SU(2)$
rotation that acts on the Pauli matrices of the initial effective
Hamiltonian. Actually this statement holds in all orders
of the perturbation theory and reads in its generality
\begin{equation}
\tilde H^{(n)}_{\text{eff}}(\phi,\theta) = g(\phi,\theta)
H^{(n)}_{\text{eff}}g^\dagger (\phi,\theta), \nonumber
\end{equation}
where $n$ is the order of the perturbation. Note that this
useful result holds not only for the $\phi$ rotations, but also
for the $\theta$ rotations which, in general, do not commute with
the collisional Hamiltonian $H^{(0)}$.

\subsection{Rotated anisotropic XY model}

We now show that the above presented Raman transitions can be employed
to obtain, for example, the anisotropic XY model. The direction of
anisotropy is determined by the phase difference of the laser
fields employed for the Raman transition. In particular consider,
as in the previous, three sites of the optical lattice in a
equilateral triangular configuration. For simplicity we assume
$J_+=J_-=J$ and $U_{\uparrow\uparrow}=U_{\downarrow
\downarrow}=U_{\uparrow \downarrow}=U$. Then the effective Hamiltonian
to the third order becomes the rotation
\begin{equation}
\tilde H_{\text{eff}}(\phi,\pi/2)=g(\phi)\tilde
H_{\text{eff}}g^\dagger(\phi), \label{rotated}
\end{equation}
where $g(\phi)=g(\phi,\theta=0)$ is a $z$-axis rotation and $\tilde
H_{\text{eff}}$ is the $\theta=\pi/2$ rotated
effective Hamiltonian around the $y$-axis given by
\begin{equation}
\tilde H_{\text{eff}}= \sum_{i=1}^3\big( A \mathbb{I}+ B\sigma_i^x
+\nu^{(1)} \sigma^x_i\sigma^x_{i+1} + \nu^{(3)} \sigma^x_i
\sigma^x_{i+1} \sigma^x_{i+2} \big), \nonumber
\end{equation}
with
\begin{equation}
A =-{3 \over 2} {J^2 \over U} -3 {J^3 \over U^2} , \,\,\, B=-2
{J^2 \over U} - {11 \over 2} {J^3 \over U^2}, \nonumber
\end{equation}
\begin{equation}
\nu^{(1)}=-{1 \over 2}{J^2 \over U} -3 {J^3 \over U^2}  ,\,\,\,
\nu^{(3)}=-{1 \over 6} {J^3 \over U^2}. \nonumber
\end{equation}
These effective couplings agree with the ones presented in
(\ref{effcouplings}). Moreover, by controlling the amplitude of the
initial standing waves that trap the atoms in their equilibrium
positions it is possible to reactivate the tunnelings $J^\uparrow$
and $J^\downarrow$. This has the effect that the overall
Hamiltonian is the sum of the two Hamiltonians, the rotated
one (\ref{rotated}) and the initial one (\ref{ham1}).

One can now check that the Hamiltonian (\ref{ham1}) is invariant
under $g(\phi)$ rotations. On the other hand, when we add the
Hamiltonians $\tilde H_0$ and the one from (\ref{ham1}) we obtain
the generalised version of the anisotropic $XY$ model where
additional third order terms are present. Hence, by turning on the
$J^\uparrow$ and $J^\downarrow$ tunnelings we can obtain the
rotated version of the anisotropic $XY$ model, where the rotation
is performed with respect to the $z$-spin axis by an angle $\phi$.
This approach provides a variety of control
parameters (e.g. the angle $\phi$ and the ratio of the couplings
of the two added Hamiltonians) and, in addition, one can have these
variables independent for each of the three directions of the two
dimensional optical lattice. Particular settings of these
structures have been proved to generate topological phenomena
\cite{Duan}, that support exotic anyonic excitations, useful for the
construction of topological memories \cite{Kitaev}. In addition,
the possibility of varying arbitrarily the control parameters of the
above Hamiltonians and, consequently, of their ground states
gives us the natural setup to study such phenomena as
geometrical phases in lattice systems. Examples will be
presented elsewhere \cite{Carollo}.

\section{Complex tunneling and topological effects}\label{complex}

Consider the case where we employ complex tunneling couplings
\cite{Zoller} in the transitions described above. This can be
performed by employing additional characteristics of the atoms like a
charge $e$, an electric moment $\vec{\mu}_e$ or a magnetic moment
$\vec{\mu}_m$, and external electromagnetic fields. As the
external fields can break time reversal symmetry, new terms of the
form $\{\sigma_{j}^x \sigma_{j+1}^y \sigma_{j+2}^z- \sigma_{j}^y
\sigma_{j+1}^x \sigma_{j+2}^z\}$ appear in the
effective Hamiltonian. In particular, the minimal coupling of the atom
with the external field can be given in general by substituting
its momentum by
\begin{equation}
\vec{p} \rightarrow \vec{p} +e \vec{A}(\vec{x}) +\vec{\mu}_m
\times \vec{E}(\vec{x}) +(\vec{\mu}_e \cdot
\vec{\nabla})\vec{A}(\vec{x}), \nonumber
\end{equation}
where $\vec{E}$ is the electric field and
$\vec{A}$ is the vector potential. All of these terms satisfy the
Gauss gauge if we demand that $\vec{\nabla} \cdot \vec{A}=0$ and
$\vec{E}(\vec{r})\propto \vec{r}/r^3$, hence they can generate
a possible phase factor for the tunneling couplings.

The first term results in the well known Aharonov-Bohm effect
\cite{Aharonov1}, while the second one is the origin of the
Aharonov-Casher effect
\cite{Aharonov2}. The first one requires that the atoms involved
are charged, which is not possible to achieve in the optical
lattice setup. On the other hand it is plausible to consider the
electric or magnetic moments of the atoms. Nevertheless the
Aharonov-Casher effect requires that the magnetic moment of the
atom moves in the field of a straight  homogeneously charged line,
the latter being technologically difficult to implement. Though,
resent experiments have been performed that generalise the
Aharonov-Casher effect partly overcoming the technological obstacles
\cite{Hinds}.
The third case involves the cyclic move of an electric moment through
a gradient of a magnetic field finally contributing the phase
\begin{equation}
\phi=\int_S(\vec{\mu}_e \cdot \vec{\nabla}) \vec{B} \cdot
d\vec{s}, \nonumber
\end{equation}
to the initial state, where $S$ is the surface enclosed by the cyclic
path of the
electric moment. For example, if $\vec{\mu}_e$ is perpendicular to
the surface $S$, taken to lie on the $x$-$y$ plane then a non-zero
phase, $\phi$, is produced if there is a non-vanishing gradient of
the magnetic field along the $z$ direction. Alternatively, if
$\vec{\mu}_e$ is
along the surface plane, then a non-zero phase is produced if the
$z$ component of the magnetic field has a non-vanishing gradient
along the direction of $\vec{\mu}_e$. Hence, it is
possible to generate a phase factor contribution to the tunneling
couplings $J=e^{i\phi}|J|$ with
\begin{equation}
\phi=\int_{\vec{x}_i}^{\vec{x}_{i+1}}(\vec{\mu}_e \cdot
\vec{\nabla})\vec{A} \cdot d \vec{x}. \nonumber
\end{equation}
Here $\vec{x}_i$ and $\vec{x}_{i+1}$ denote the positions of the
lattice sites connected by the tunneling coupling $J$.

In order to isolate the new terms that appear in the case of
complex tunneling couplings we should restrict to purely imaginary
ones, i.e. $J_{j}^{\sigma}= \pm i |J_{j}^{\sigma}|$. Then the
effective Hamiltonian (\ref{effective}) becomes
\begin{equation}
\begin{split}
H_{\text{eff}}=&\sum_i \Big[A \mathbb{I} + B \sigma^z_i
+\\&\tau^{(1)}\sigma^z_i\sigma^z_{i+1}
+\tau^{(2)}(\sigma^x_i\sigma^x_{i+1}+ \sigma^y_i\sigma^y_{i+1})+ \\ \\
&\tau^{(3)}(\sigma^x_i\sigma^y_{i+1}- \sigma^y_i\sigma^x_{i+1})+
\tau^{(4)}\epsilon_{lmn}\sigma^l_i \sigma^m_{i+1} \sigma^n_{i+2}
\Big],
\end{split}
\label{complexboson}
\end{equation}
where $\epsilon_{lmn}$ with
$\{l,m,n\}=\{x,y,z\}$ denotes the total antisymmetric tensor in
three dimensions and summation over the indices $l,m,n$ is
implied. The couplings appearing in (\ref{complexboson}) are given
in the bosonic case by
\begin{equation}
\begin{split}
&A= {{J^\uparrow}^2 \over U_{\uparrow\uparrow}} +{{J^\downarrow}^2
  \over U_{\downarrow\downarrow}} +{ {J^\uparrow}^2+{J^\downarrow}^2 \over 2
  U_{\uparrow \downarrow}},\,\,\,
B= 2{{J^\uparrow}^2 \over U_{\uparrow\uparrow}} -2{{J^\downarrow}^2
  \over U_{\downarrow\downarrow}},
\\ & \tau^{(1)} = {{J^\uparrow}^2 \over U_{\uparrow\uparrow}} +{{J^\downarrow}^2
  \over U_{\downarrow\downarrow}} -{ {J^\uparrow}^2+{J^\downarrow}^2 \over 2
  U_{\uparrow \downarrow}},\,\,\,
\tau^{(2)}={J^\uparrow J^\downarrow \over
  U_{\uparrow\downarrow}},\\
&\tau^{(3)}= i{{J^\uparrow}^2 J^\downarrow \over
  U_{\uparrow\uparrow}}\Big( {1 \over 2 U_{\uparrow\uparrow}} + {1
  \over U_{\uparrow\downarrow}} \Big) + (\uparrow \leftrightarrow
  \downarrow),\,\,\,
\\&
\tau^{(4)}= i{{J^\uparrow}^2 J^\downarrow \over
  U_{\uparrow\uparrow}}\Big( {1 \over 2 U_{\uparrow\uparrow}} + {1
  \over U_{\uparrow\downarrow}} \Big) - (\uparrow \leftrightarrow
  \downarrow), \nonumber
\end{split}
\end{equation}
and in the fermionic case by
\begin{equation}
\begin{split}
&A=-\tau^{(1)}= {{J^\uparrow}^2 +{J^\downarrow}^2 \over 2 U },\,\,\,\, B=\tau^{(3)}=0
, \\
&\tau^{(2)}=-{{J^\uparrow}{J^\downarrow} \over U },\,\,\,\,
\tau^{(4)}= i {{J^\uparrow}^2 J^\downarrow
-{J^\downarrow}^2 J^\uparrow \over 2U^2}. \nonumber
\end{split}
\end{equation}
By taking $U_{\uparrow \downarrow} \rightarrow \infty$, $U_{\uparrow 
\uparrow}=-U_{\downarrow \downarrow}=-U$, $J^\uparrow=-J$ and
$J^\downarrow=J$, one can set, in the bosonic case, with the aid of
Feshbach resonances and compensating Zeeman terms, all the couplings
to be zero apart from $\tau^{(4)}$. Hence, the effective Hamiltonian
reduces to
\begin{equation}
H_{\text{eff}}= \tau^{(4)} \sum_{\langle ijk
  \rangle}\vec{\sigma}_i\cdot \vec{\sigma}_j
\times \vec{\sigma}_k, \label{chiral}
\end{equation}
with $\vec{\sigma}=(\sigma^x,\sigma^y,\sigma^z)$ and
$\tau^{(4)}=|J|^3/U^2$. Remarkably, with this physical proposal, the
interaction term (\ref{chiral}) can be isolated, especially from the
Zeeman terms that are predominant in equivalent solid state
implementations. This interaction term is also known in the literature
as the {\em chirality operator} \cite{Wen}. It breaks time reversal
symmetry of the system, a consequence of the externally applied field,
by effectively splitting the degeneracy of the ground state into two
orthogonal sectors, namely ``$+$'' and ``$-$'', related by time
reversal, $T$. These sectors are
uniquely described by the eigenstates of $H_{\text{eff}}$ at the
sites of one triangle. The lowest energy sector with eigenenergy
$E_+=-2\sqrt{3} \tau^{(4)}$ is given by
\begin{equation}
\begin{split}
&|\Psi^+_{1/2} \rangle = {1 \over \sqrt{3}} \big(|\uparrow
\uparrow \downarrow \rangle + \omega |\uparrow \downarrow \uparrow
\rangle + \omega^2 |\downarrow \uparrow
\uparrow \rangle \big) \\
& |\Psi^+_{-1/2} \rangle =- {1 \over \sqrt{3}} \big(|\downarrow
\downarrow \uparrow \rangle + \omega |\downarrow \uparrow
\downarrow \rangle + \omega^2 |\uparrow \downarrow \downarrow
\rangle \big) \label{states}
\end{split}
\end{equation}
The excited sector, $|\Psi^-_{\pm1/2} \rangle$, represents counter
propagation with eigenvalue $E_-=2\sqrt{3}\tau^{(4)}$ and it is obtained from
(\ref{states}) by complex conjugation \cite{Wen,Rokhsar,Sen}. We would
like to point out that, to the best of
our knowledge, this is the first physical proposal where this
interaction term can be isolated, especially from the Zeeman terms
that are predominant in equivalent solid state implementations.
Alternative models employing cold atom technology for the
generation of topologically non-trivial ground states are given in
\cite{Duan,Ruostekoski}.

\section{One- and Two-dimensional models} \label{onedim}

It is also possible to employ the three-spin interactions that we
studied extensively in the previous sections for the construction
of extended one and two dimensional systems. The two dimensional
generalisation is rather straightforward as the triangular system
we considered is already defined on the plane. Hence, all the
interactions considered so far can be generalised for the case of
a two dimensional lattice where the summation runs through all the
lattice sites with each site having six neighbours.

\begin{center}
\begin{figure}[ht]
\resizebox{!}{2.5cm}
{\includegraphics{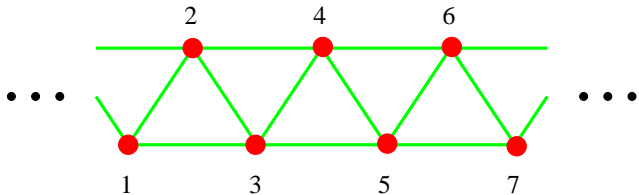} }
\caption{\label{chain} The one dimensional chain constructed out of
    equilateral triangles. Each triangle experiences the three-spin
    interactions presented in the previous.}
\end{figure}
\end{center}

The construction of the one dimensional model is more involving.
In particular, we now consider a whole chain of triangles in a zig-zag
one dimensional pattern as shown in Fig. \ref{chain}. In principle this
configuration can extend our model from the triangle to a chain.
Nevertheless, a careful consideration of the two spin interactions
shows that terms of the form $\sigma^z_i \sigma^z_{i+2}$
appear in the effective Hamiltonian, due to the triangular setting
(see Fig. \ref{chain}). Such
Hamiltonian terms involving nearest and next-to-nearest neighbour
interactions are of interest in their own right
\cite{Sachdev,Sachdev1} but will not be address here. It is also
possible to introduce a longitudinal optical lattice
with half of the initial wavelength, and an appropriate amplitude
such that it cancels exactly those interactions generating, finally,
chains with only neighbouring couplings.
In a similar fashion it is possible to avoid generation of
terms of the form $\sigma^x_i \sigma^x_{i+2} + \sigma^y_i
\sigma^y_{i+2}$ by deactivating the longitudinal tunneling coupling in
one of the modes, e.g. the $\uparrow$ mode, which deactivates the
corresponding exchange interaction.

As we are particularly interested in three-spin interactions we
would like to isolate the chain term $\sum_i (\sigma^x_i
\sigma^z_{i+1} \sigma^x_{i+2}+\sigma^y_i \sigma^z_{i+1}
\sigma^y_{i+2})$ from the $\lambda^{(4)}$ term of Hamiltonian
(\ref{ham1}). This term includes, in addition, all the possible
triangular permutations. To achieve that we could deactivate the
non-longitudinal tunneling for one of the two modes, e.g. the one
that traps the $\uparrow$ atoms. The interaction $\sigma^z_i
\sigma^z_{i+1} \sigma^z_{i+2}$ is homogeneous, hence it does not
pose such problem when it is extended to the one dimensional
ladder. With the above procedures we can finally obtain a chain
Hamiltonian as in (\ref{ham1}) where the summation runs up to the
total number $N$ of the sites.

\section{Conclusions}
\label{conclusions}

In this paper we presented a variety of different spin
interactions that can be generated by a system of ultra-cold atoms
superposed by optical lattices and initiated in the Mott insulator
phase. In particular, we have been interested in the simulation
and study of various three-spin interactions conveniently obtained
in a lattice with equilateral triangular structure. They appear by
a perturbation expansion to the third order with respect to the
tunneling transitions of the atoms when the dominant interaction
is the collisions of atoms within the same site. Among the models
presented here we specifically considered the $\sigma^z_i
\sigma^z_{i+1} \sigma^z_{i+2}$ interaction, a third order
generalisation of the rotated inhomogeneous $XY$ model, as well as
interactions that explicitly break chiral symmetry. These models
can exhibit degeneracy in their ground states and undergo a
variety of quantum phase transitions that can also be viewed as
phases of the initial Mott insulator.

It is possible to employ quantum simulation techniques
\cite{Jane}, in a similar fashion as for two spin Hamiltonians, to
generate effective three-spin interactions that are not possible
to obtain straightforwardly from the optical lattice system.
Hence, a variety of additional Hamiltonians can be obtained by
considering manipulations of the above three-spin interactions
with the application of appropriate instantaneous one or two spin
transformations. The possibility to externally control most of the
parameters of the effective Hamiltonians at will reenters our
model as a unique laboratory to study the relationship among
exotic systems such as chiral spin systems, fractional quantum
Hall systems or systems that exhibit high-$T_c$ superconductivity
\cite{Wen,Laughlin}. In addition, suitable applications have been
presented within the realm of quantum computation \cite{Pachos03}
where three-qubit gates can be straightforwardly generated from
the three-spin interactions. Furthermore, unique properties
related with the criticality behaviour of the chain with three-spin
interactions has been analysed in \cite{Pachos04} where the
two-point correlations, used traditionally to describe the
criticality of a chain, seem to fail to identify long quantum
correlations, suitably expressed by particular entanglement
measures \cite{Verstraete PC 03}.

In conclusion, we have presented a physical model that can
efficiently simulate a variety of three-spin interactions. The
employed optical lattice techniques give the possibility to
externally manipulate and control the couplings of the
interactions. The effect of these terms will eventually be
significant with the improvement of the experimental techniques.
Importantly, the three-spin interactions can be isolated from
two-spin ones or from possible Zeeman terms that are always
present in the corresponding spin systems. This makes the further
study of their properties an important task for future work.

\acknowledgements

We thank Derek Lee for inspiring conversations and Almut
Beige, Alastair Kay and Martin Plenio for
critical reading of the manuscript. This work was
supported by a Royal Society University Research Fellowship and by
the Spanish grant MECD AP2001-1676. E.R. thanks the QI
group at DAMTP for their hospitality, where part of this work was
done.

\appendix

\section{Perturbation theory}

Consider the case of two species of atoms trapped in optical
potentials forming a triangular configuration subject to the
Hamiltonian given by (\ref{ham}). For simplicity define the
diagonal free Hamiltonian to be given by $H^{(0)}_{ij}= E_i
\delta_{ij}$, where $E_i$ is either zero or proportional to
$U_{aa}$, $U_{bb}$ or $U_{ab}$. As we have already mentioned we
consider the case where tunneling couplings are much smaller than
the collisional ones $J\ll U $. Then the evolution of the system
is dominated by the term $H^{(0)}$. In fact, when we start from a
configuration of one atom per lattice site, denoted by the
subspace $M$ of configurations, and activate small tunneling
couplings, the change of atom number per site would be
energetically unfavourable and is hence adiabatically eliminated.

To see this analytically we employ the interaction picture with
respect to the Hamiltonian $H^{(0)}$ obtaining
\begin{equation}
{H_I}_{ij}(t)=V_{ij} \exp \Big[ {\rm i}(E_i -E_j)t/\hbar \Big].
\end{equation}
The evolution operator in the interaction picture is given by
the time ordered formula
\begin{eqnarray}
&&{\cal U}_I(t,0) \equiv
\text{T} \exp \Big[ -\frac{{\rm i}}{\hbar}
  \int _0^t H_I (t')dt'\Big]
\no \no &&=\mathbb{I} - {{\rm i} \over \hbar} \int _0^t H_I
(t')dt' - {1 \over \hbar^2} \int_0^t dt' \int_0^{t'}
dt''H_I(t')H_I(t'') \no \no &&+ {{\rm i} \over \hbar ^3} \int_0^t
dt' \int_0^{t'} dt'' \int_0^{t''} dt'''H_I(t') H_I(t'') H_I(t''')
\no \no &&+ {\cal O}((Jt)^4). \label{evol}
\end{eqnarray}
Higher orders are negligible as long as times $t$ are considered for
which $Jt$ remains sufficiently small, while
$Ut$ is large enough to avoid the accumulation of population
outside the subspace $M$. The latter condition is necessary to
exempt fast rotating phase factors appearing when performing the above
time integrals. These phase factors are of the form $e^{i \omega t}
-1$ and
\begin{eqnarray}
&& \lim_{t \rightarrow \infty}\big( e^{i \omega t} -1\big) =\lim_{t
    \rightarrow \infty}\big( -2\sin^2 {\omega t \over 2} +i \sin
    \omega t \big)
\no \no && = (-\omega^2 t \pi +i 2 \pi \omega) \delta(\omega), \nonumber
\end{eqnarray}
which is zero for $\omega\propto E_i-E_j \neq 0$. These conditions
are in agreement with the previous demands that $Jt$ is very
small while $Ut$ is relatively large. Hence, we can
directly calculate each term of the expansion (\ref{evol}) without
having to take into account fast rotating terms.

The effective Hamiltonian $H_{\text{eff}}$ that corresponds to this
evolution can be obtained by the term proportional to time $t$ in the
expansion of the evolution operator, i.e.
\begin{equation}
{\cal U}_I(t,0) = \mathbb{I}-{{\rm i} \over \hbar }
H_{\text{eff}} t+ {\cal O}(t^2) \nonumber
\end{equation}
Consider now the second term on the right hand side of (\ref{evol}).
This term gives no evolution within the subspace $M$ of states as the
tunneling Hamiltonian term $V$ moves you necessarily out of the $M$
configurations. The third term gives (see \cite{Kukl})
\begin{equation}
({H^{(2)}_{\text{eff}}})_{\alpha \beta}=-\sum_\gamma {V_{\alpha
\gamma} V_{\gamma \beta} \over E_\gamma}, \label{second}
\end{equation}
where $\alpha$ and $\beta$ indicate states in $M$, $\gamma$
indicates states out of $M$ and $E$ are the eigenstates of
$H^{(0)}$, where we have used $E_\alpha=E_\beta=0$. This gives the
usual second order effective Hamiltonian presented in detail in
\cite{Kukl,Duan}. Consider now three sites and the effect of the
third term in Eqn. (\ref{evol}). Finally, we obtain the
effective Hamiltonian with matrix elements
\begin{equation}
({H^{(3)}_{\text{eff}}})_{\alpha \beta}= \sum_{\gamma \delta}
   {V_{\alpha \gamma} V_{\gamma \delta} V_{\delta \beta} \over
     E_\gamma E_\delta} .
\label{third}
\end{equation}

With the formulae (\ref{second}) and (\ref{third}) one can perform the
perturbation up to the
third order and find the desired three-spin interactions (\ref{effective}). In
practise the evaluation of the terms that contribute to the
three-spin Hamiltonian is quite simple. The states corresponding
to $\gamma$ and $ \delta$ include sites with two or three atoms of
the same or of different species. Hence, $E_\gamma,E_\delta
\propto U_{\sigma\sigma'}$. Next you need to consider the
different evolutions of the form
$\alpha\rightarrow\gamma\rightarrow\delta\rightarrow\beta$, that
populations undertake. The tunneling couplings $J^\sigma$ are
determined by each of these transitions and an appropriate
coefficient is obtained in the case of the bosonic generation or
annihilation of two atoms of the same species in one site.

\section{Adiabatic Elimination}

As an alternative procedure it is possible to eliminate the fast
oscillating term without
performing the perturbative expansion. This elimination is related
with the adiabatic elimination of the states with two or more atoms
per lattice site that are separated from the states with one atom per
lattice site (configurations in $M$) by a large energy gap
proportional to $U_{\sigma \sigma'}$. In fact if we set a
decomposition of the three site in
terms of basis states of the form $|i_1j_1;i_2,j_2;i_3,j_3\rangle$
where $1,2,3$ denote the site, and $i_k$ and $j_k$ denote the number of
atoms of species $\uparrow$ and $\downarrow$ respectively in site $k$ we can write the
general state of the three sites as
\begin{equation}
|\Psi(t)\rangle= \sum_{i_k,j_k} c^{i_1 i_2 i_3}_{j_1j_2
  j_3}(t)|i_1j_1;i_2,j_2;i_3,j_3\rangle .
\nonumber
\end{equation}
By employing the Schr\"odinger equation we can obtain the
time-differential equations of the coefficients $c^{i_k}_{j_k}$ of the
form
\begin{equation}
i \hbar \dot c^{i_k}_{j_k} = \sum _{i_k',j_k'} H^{i_k j_k}_{i_{k'}
  j_{k'}} c^{i_k'}_{j_k'} ,
\label{schr}
\end{equation}
where $H^{i_k j_k}_{i_{k'} j_{k'}} = \langle c^{i_{k}}_{j_{k}}| H
|c^{i_{k'}}_{j_{k'}}\rangle$. It is easy to verify that the
elements of $H$ with indexes $(i_k j_k)$ corresponding to states
that do not belong to $M$ include fast rotating phases and, hence,
they are zero, i.e. for those states $\dot c^{i_k}_{j_k}=0$. This
provides a set of linear equations of the form $\sum _{i_k',j_k'}
H^{i_k j_k}_{i_{k'} j_{k'}} c^{i_k'}_{j_k'}=0$ that can be solved,
in principle, explicitly. In our case, Eqn. (\ref{schr}) has
overall 56 equations resulting from the Schr\"odinger
equation with 48 reduced to a linear system of coupled
algebraic equations. This set can be solved by a computer and then
expanded in terms of small couplings  $J\ll U$ obtaining an
alternative verification of the previous perturbative expansion.

\end{document}